# DESIGN OF THE ESS RFQs AND CHOPPING LINE

R. Duperrier, R. Ferdinand, P. Gros, J-M. Lagniel, N. Pichoff, D. Uriot
CEA-Saclay, DSM-DAPNIA-SEA

*Abstract*

The chopping line is a critical part of the ESS linac in term of technical realisation of the choppers and preservation of the beam qualities. A new optimised design of the ESS RFQs and chopping lines is reported. The beam dynamics has been optimised with H$^-$ beam currents up to 100-mA to have safety margin with respect to the ESS goals. The first RFQ transmits almost 99.7% of the beam up to 2 MeV. The line with two choppers allows a perfect chopping between 2 bunches. The second RFQ accelerates the particles up to 5 MeV with a transmission close to 100%.

## 1 INTRODUCTION

The European Spallation Source (ESS) reference design is described in ref. [1]. The accelerator is designed to provide a proton beam power of 5 MW at a repetition rate of 50 Hz. It comprises a 1.334 GeV H$^-$ linac (~ 10% duty cycle) and two accumulator rings. The proposed lay-out of the linac has two front ends, each one made up of an 70 mA peak H$^-$ ion source, a low energy beam transport, a first RFQ, a Medium Energy Beam Transport (MEBT) with the choppers and a second RFQ. The bunch funnelling is done at 5 MeV and a 350 MHz Drift Tube Linac (DTL) accelerates the beam up to 70 MeV. In this reference design a 700 MHz normal conducting Coupled Cavity Linac (CCL) further accelerates the H- beam to the final energy.

## 2 DESCRIPTION

In high power proton accelerators for projects such as ESS or the multi-user facility (CONCERT project [2]), a chopped beam is needed to reduce particle losses at injection in the accumulator rings. The chopping system must be also used to create gaps between the batches sent to the different users by fast switching magnets. The choice of 2 MeV as chopping energy results of a compromise between space-charge induced debunching and emittance growth at low energy, chopper feasibility problems increasing with the energy and the necessity to stay below the first radio-activation threshold of copper (2.16 MeV).

The H$^-$ source is pulsed with a rise/falling time of about 10 µs [3]. The beam is then pre-chopped in the LEBT with a rising/falling time close to 1 µs. The optimisation has been done with the beam injected at 95 keV in the first RFQ. Experiences gained on high power linac [4,5] show that safety margins on both beam current *and* beam emittance has to be taken to obtain a robust design with nominal parameters. For that purpose, 100 mA H$^-$ beam with 0.25 π.mm.mrad transverse *rms* norm emittances was the reference for the study presented here. The goal was to preserve the beam quality while the chopping occurs between 2 rf bunches in about 2 ns. Multiparticle codes have been used to optimise the whole line (RFQ1 – MEBT – RFQ2) to ensure realistic calculations of both transmission and emittance growth.

### 2.1 RFQ1

The first RFQ is designed to accelerate the beam from 95 keV to 2 MeV. The beam dynamics is computed using both PARMTEQM (z-code) and TOUTATIS, a more sophisticated t-code [6]. The optimisation procedure is to design an initial RFQ for the full energy range (95 keV up to 5 MeV) and then to cut it in two parts. The maximum electric field is maintained below 1.7 Kp (31.3 MV/m) in order to avoid sparks in the cavity. The main RFQ1 parameters are specified in Table 1.

Table 1: RFQs specifications and results

| *Parameter* | *RFQ1 Value* | *RFQ2 value* |
| --- | --- | --- |
| Input energy | 95 keV | 2.0 MeV |
| Output energy | 2.0 MeV | 5 MeV |
| Input current | 100 mA | 97.1 mA |
| Input emit. | 0.25 π.mm.mrad | 0.29 π.mm.mrad |
| Length | 5 m | 3 m |
| Number of cells | 527 | 83 |
| Min. aperture a | 3.52-4.13 mm | 3.71-3.73 mm |
| Modulation | 1-1.59 | 1.59-1.75 |
| Vane voltage | 87.3-117.7 kV | 117.7-122.8 kV |
| Output emit. | 0.26 π.mm.mrad | 0.3 π.mm.mrad |
| Transmission | 99.7 % | 99.97 % |

A transition cell is included between the last accelerating cell and the fringe field [7] and the length of this fringe field is adjusted to simplify the MEBT design. The length of the RFQ1 cavity is mainly imposed by the slow adiabatic bunching process needed to reach a high capture efficiency as in the IPHI project [4]. The RF segmented RFQ concept is kept from previous project [4,5] for the construction of the ESS RFQ cavities.

### 2.2 MEBT

The aim of the Medium Energy Beam Transport line is to match the beam into the second RFQ minimising emittance growth and halo formation. Less than 0.01% of the chopped beam must enter the second RFQ, and the transmission of the non-chopped beam must be higher than 95 %. The line is made up of 10 quadrupoles, 3 bunchers and 2 choppers. The total length (2.1m) is kept as short as possible to minimise the emittance growth. The 2 choppers are installed inside the quadrupoles.

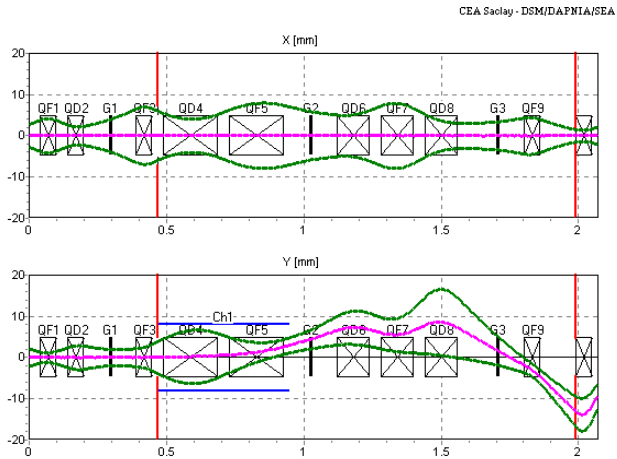

Figure 1: Medium energy beam line.
Top : normal transport in the X plane.
Bottom : Y plane with the choppers (blue).
The 2 diaphragms are shown in red.

The first diaphragm prevents damages on the choppers. The second one collects the chopped beam and cleans the non-chopped beam before RFQ2 (about 7 kW of stopped particles is expected from the chopped beam). They are sectored diaphragm to help in the tuning process.

| Chopper plate voltage | Line Transmission |
|---|---|
| 800 V | 0.01 % |
| 600 V | 0.16% |
| 400 V | 11.0 % |
| 0 V | 97.2 % |

Table 2: transmission vs the dynamic plate voltage

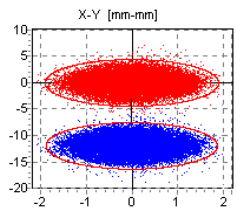

Figure 2 : Chopped and unchopped beam

| Emittance growth | X | Y | Z |
|---|---|---|---|
| 100 mA | + 11 % | + 13 % | + 6 % |
| 70 mA | + 6 % | + 9 % | + 1 % |

Table 3: Emittance increase through the MEBT.

Due to the non-linearity of the buncher fields and the space-charge induced radial–longitudinal coupling, emittance growth are minimum when the bunch transverse and longitudinal sizes in the bunchers are small. The line has been optimised to avoid these effects leading to the emittance growth given in Table 3 and a transmission of 97.2 %.

## 2.3 Choppers

**Chopper requirements -** The chopper tricky task is to clear off an intact number of rf bunches with drastic rise/fall times (Table 4).

| Electric length | 2 x 240 mm | or 1 x 480 mm |
|---|---|---|
| Gap | 16 mm | Min |
| Pulser voltage | +/- 950 V | Min |
| Field efficiency | 84 % | |
| Chopping time | 2 × 600 μs | 100 μs spaced out |
| Chopping frequency | 360 ns off, 240 ns on, 50 Hz | |
| Duty factor | 5-65 % | |
| Rise/fall time | < 2 ns | 2-98 % |

Table 4: chopper main requirements

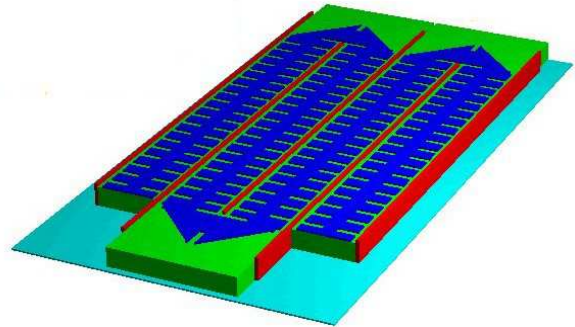

Figure 3 : 4 tracks 3D sections of the MAFIA model

**Technologic choices -** A micro-strip line meander structure (Figure 3), like the one chosen for the SNS design [8] [9], seems to suit this type of fast beam chopping. The total strip length has to be limited to avoid pulse distortion. The choice of a notched line allows large strips width on thin laminate keeping a 50 ohms characteristic impedance [10]. The rise and fall time requirement in the ns range implies a quasi TEM mode for the signal wave propagation.

**Simulations and calculations -** 2D simulations [11] give:
- The capacitance per unit length of the micro-strip line using a static field solver.
- The self inductance per unit length using a transient solver which takes into account the skin effect.

The self inductance increase due to notches is empirically calculated from previous slit studies in micro-strip lines [12]. The chamfered extremities can be predetermined from charts [13]. The characteristic impedance of the line and the signal phase velocity can be deducted from these calculations.

Figure 4 shows a coloured contour density of the electrical field and equipotential lines. The field ripple is less than 5% near the beam axe. Beam dynamics calculations done using multiparticle codes do not show evident effects resulting from this ripple.

MAFIA code [14] allows full 3D simulations in the temporal domain. It confirms the previous calculations within a 5% error margin. New calculations with mesh refinement will be performed soon.

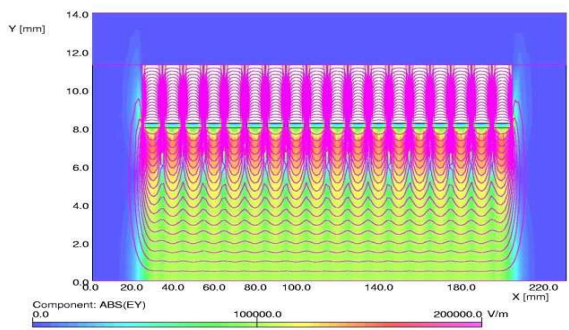

Figure 4: 2D electric field density
(18 tracks cross-section)

Moreover Pspice electrical simulations [15] dealing with "lossy" line fitted with appropriate coefficients let us assert that we can reach the desired rise and fall times with a single 50 cm long chopper. Nevertheless, two 24 cm long choppers were used for the simulations. Table 5 summarises the geometrical and electrical values to reach the chopping requirements.

| Microstrip width | 8 mm | |
|---|---|---|
| Tracks period | 10 mm | |
| Separator thickness | 500 µm | 250 µm overhanging |
| Meander line width | 78.2 mm | 60.2 mm straight |
| Laminate thickness | 3.04 mm | Rogers RT 6002 |
| Notches period | 3.6 mm | |
| Notch depth / width | 3 / 1 mm | |
| Lineic inductance | 335 nH/m | |
| Lineic capacitance | 134 pF/m | |
| Characteristic impedance | 50 ohms | |
| Phase velocity | 150 mm/ns | 0.5*c |

Table 5 : calculated main electrical characteristics.

| Total line length | 2.87 m |
|---|---|
| Plate length | 388 mm 38 tracks |
| Plate width | 91 mm |
| Gap stroke | 10-35 mm |
| Line thickness | 70 µm |
| Global tolerance | 5/100 mm |

Table 6 : prototype main characteristics

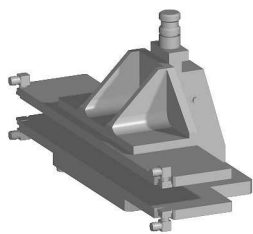

Figure 5 : 3D prototype realistic view

**Prototype design -** A prototype (Figure 5) has been recently launched and will be ready for tests and measurements for the end of summer 2000. It is a full 2 plates structure with limited length and line thickness (Table 6), a limitation due to the use of a standard microwave laminate (two Rogers RT 6002 12"x18" plates 1.52 mm thick pasted together). A micrometric slide allows the stroke adjustment to measure the impedance variation according to the gap range. A TNC connector terminates each line end.

## 2.4 RFQ2

The second RFQ brings the bunched beam from 2 MeV to 5 MeV. An inverse transition cell is added between the short matching section and the beginning of the modulated vanes in order to preserve the beam emittances. The 5 MeV final output energy allows the construction of the first drift tubes of the DTL using EM quadrupoles as demonstrated by the IPHI R&D programme. Again, 100 000 particles have been transported from RFQ1 to RFQ2 through the MEBT to ensure realistic simulations. The main RFQ2 parameters are shown in Table 1.

## 3 CONCLUSION

The present design of the two 352 MHz RFQs and the chopping line allows a perfect chopping between two bunches without beam characteristics degradations. A prototype is in construction and will be tested soon to confirm the calculated performance of the chopper.

## 4 ACKNOWLEDGEMENTS

The authors would like to thanks Sergey S. Kurennoy (LANL) for his helpful advises and information.